\documentclass[prd,aps,nofootinbib,showpacs,preprintnumbers,showkeys]
{revtex4}
\usepackage{graphicx,epsf,amsmath,amsfonts,amssymb,amsbsy}
\newcommand{\ds}{\displaystyle}
\newcommand{\vev}[1]{\langle#1\rangle}
\newcommand{\mat}{\left ( \begin{array}}
\newcommand{\emat}{\end{array} \right )}
\newcommand{\vect}{\left ( \begin{array}{c}}
\newcommand{\evect}{\end{array} \right )}

\preprint{HU-EP-07/18}

\begin{document}
\title{Mass spectrum of diquarks and mesons in the color--flavor
locked phase of dense quark matter}
\author{D.~Ebert$^{\ast}$, K.~G.~Klimenko$^{\dagger,\ddagger}$, V.L.
Yudichev$^{\flat}$}
\affiliation{$^{\ast}$ Institut f\"ur Physik,
Humboldt-Universit\"at zu Berlin, 12489 Berlin, Germany}
\affiliation{$^{\dagger}$ Institute
for High Energy Physics, 142281, Protvino, Moscow Region, Russia}
\affiliation{$^{\ddagger}$ Dubna University (Protvino
branch), 142281, Protvino, Moscow Region, Russia}
\affiliation{$^{\flat}$ Joint Institute for Nuclear Research 141980,
Dubna, Moscow Region, Russia}

\begin{abstract}
The spectrum of  meson and diquark excitations of dense quark matter 
is considered in the framework of the Nambu -- Jona-Lasinio model
with three types of massless quarks in the presense of a quark number
chemical potential $\mu$. We investigate the effective action
of meson- and diquark fields both at sufficiently large values of
$\mu>\mu_c\approx 330$ MeV, where the color--flavor locked (CFL)
phase is realized, and in the chirally broken phase of quark matter 
($\mu<\mu_c$). In the last case all nine pseudoscalar mesons are
Nambu -- Goldstone (NG) bosons, whereas the mass of the scalar meson
nonet is twice the dynamical quark mass. In the chirally broken phase
the pseudoscalar diquarks are not allowed to
exist as stable particles, but the scalar diquarks might be stable
only at a rather strong interaction in the diquark channel. In the
case of the CFL phase, all NG bosons of the model are realized as
scalar and pseudoscalar diquarks. Moreover, it turns out that 
massive diquark excitations are unstable for this phase. 
In particular, for the scalar and pseudoscalar octets of diquark
resonances a mass value around 230 MeV was found numerically.
In contrast, mesons are stable particles in the CFL phase. Their
masses lie in the interval 400$\div$500 MeV for not too large values
of $\mu>\mu_c$.
\end{abstract}


\maketitle

\section{Introduction}

It is well-known that at asymptotically high baryon densities 
the ground state of massless three-flavor QCD corresponds to the
so-called color -- flavor locked (CFL) phase \cite{alford1,alford2}. 
In this phase quarks of all three flavors as well as three colors
undergo pairing near the Fermi surface due to the attractive
one-gluon exchange potential. 
The properties of different collective modes, including Nambu --
Goldstone (NG) bosons, of the CFL phase were studied already in the
framework of weak-coupling QCD \cite{rho}.
At intermediate baryon densities, related to compact star physics, 
weak-coupling expansion of QCD is not applicable, so the description
of color superconductivity, including the CFL phase, is more adequate
in the framework of effective theories for the low energy QCD
region. In particular, since massless excitations might play an
important role in different transport phenomena, such as cooling
processes of neutron stars etc,
different chiral type effective theories
for the pseudoscalar NG bosons of the CFL phase are usually used 
(see, e.g., \cite{casalbuoni,andersen}).

Another effective theory approach 
is based on the Nambu -- Jona-Lasinio (NJL) models. Since any NJL
model contains the microscopic quark degrees of freedom, it is 
especially convenient for the investigation of dynamical processes in
dense baryonic matter. In particular, in the three-flavor NJL model
the CFL effect was already considered, e.g., in \cite{nebauer}, (see
also the review \cite{buballa}), where some aspects of the phase
structure of dense quark matter were discussed, including the
influence of the $s$-quark bare mass, color- and electric charge
neutrality conditions, external magnetic field, etc. In addition, in
\cite{reddy,ek} the properties and structure of NG bosons of the CFL
phase were considered in the framework of NJL models.

One of the most noticeable differences between color
superconductivity phenomena with three and two quark species is that
the CFL effect is characterized by a hierarchy of energy scales. As
it was established in different approaches quoted above, at the
lowest scale lie NG bosons, which dominate in all physical processes
with energy smaller, than the superconducting gap $\Delta$. 
Evident contributors at higher energy scales are quark
quasiparticles, which in the CFL phase have an energy greater than
$\Delta$. However, up to now we know much less about other
excitations, whose energy and mass are of the order of $\Delta$ in
magnitude. Among these particles are ordinary scalar and pseudoscalar
mesons, massive diquarks etc, i.e. particles which might play an
essencial role in dynamical processes of the CFL phase. In contrast,
the properties of mesons and diquarks, surrounded by color
superconducting quark matter, were already discussed in the framework
of the two-flavor NJL model \cite{bekvy,eky1,eky2,he,hashimoto}.

In the present paper we are going to study just this type of
excitations of the CFL ground state, i.e. mesons and massive
diquarks, in the framework of the massless three-flavor NJL model. 
In our previous paper \cite{ek} we have, in particular, 
obtained the equations for both scalar and pseudoscalar diquark
masses in the CFL phase of the NJL model. 
There, our consideration was based on the effective action, which is
a generating functional for one-particle irreducible Green functions.
Now, using the same technique, we perform a numerical investigation
of diquark masses, as well as of the masses of scalar and
pseudoscalar mesons, vs the chemical potential in the CFL phase. 
Moreover, the octet and singlet structure is established for massive
mesons and diquarks (both scalar and pseudoscalar) in the CFL phase. 
In addition, the masses of diquarks and mesons in the chirally broken
quark matter phase are also investigated.

\section{NJL model and its effective action}

Let us consider the following NJL model with three massless quark
flavors
\begin{eqnarray}
&&  L=\bar q\Big [\gamma^\nu i\partial_\nu+  \mu\gamma^0\Big
]q+ G_1\sum_{a=0}^8\Big [(\bar
q\tau_aq)^2+(\bar qi\gamma^5\tau_a q)^2\Big ]+\nonumber\\
&+&G_2\!\!\!\sum_{A=2,5,7}\sum_{A'=2,5,7}\Big\{
[\bar q^Ci\gamma^5\tau_A\lambda_{A'}q]
[\bar qi\gamma^5\tau_A\lambda_{A'} q^C]
+[\bar q^C\tau_A\lambda_{A'}q]
[\bar q\tau_A\lambda_{A'} q^C]\Big\}.
   \label{1}
\end{eqnarray}
In (\ref{1}), $\mu\geq 0$ is the quark number chemical potential
which is the same for all quark flavors, $q^C=C\bar q^t$, $\bar
q^C=q^t C$ are charge-conjugated spinors, and
$C=i\gamma^2\gamma^0$ is the charge conjugation matrix (the symbol
$t$ denotes the transposition operation). The quark field
$q\equiv q_{i\alpha}$ is a flavor and color triplet as
well as a four-component Dirac spinor, where $i,\alpha=1,2,3$.
(Roman and Greek indices refer to flavor and color
indices, respectively; spinor indices are omitted.) Furthermore,
we use the notations  $\tau_a,\lambda_a$ for Gell-Mann matrices in
the flavor and color space, respectively ($a=1, ...,8)$; $\tau_0
=\sqrt{\frac{2}{3}}\cdot\bf 1_f$ is proportional to the unit
matrix in the flavor space. Clearly, the Lagrangian (\ref{1}) as a
whole is invariant under transformations from the color group
SU(3)$_c$. In addition, it is symmetric under the chiral group
SU(3)$_L\times$SU(3)$_R$ (chiral transformations act on the flavor
indices of quark fields only) as well as under the baryon-number
conservation group U(1)$_B$ and the axial group U(1)$_A$. 
\footnote{In a more realistic case, the additional
`t Hooft six-quark interaction term should be taken into account in
order to break the axial U(1)$_A$ symmetry \cite{nebauer}. However,
in the present consideration we omit the `t Hooft term, for
simplicity.} 
In all numerical 
calculations below, we use the following values of the model
parameters (see, e.g., ref. \cite{buballa}): $\Lambda=602.3$ MeV,
$G_1\Lambda^2=2.319$ and $G_2=3G_1/4$, where $\Lambda$ is an
ultraviolet cutoff parameter in the three-dimensional momentum space. 

The linearized version of the Lagrangian (\ref{1}) contains
collective bosonic fields $\sigma_a (x),\pi_a
(x),\Delta^{s}_{AA'}(x),\Delta^{p}_{AA'}(x)$ and looks like
\begin{eqnarray}
\tilde L\ds &=&\bar q\Big [\gamma^\nu i\partial_\nu +\mu\gamma^0
-\sigma_a\tau_a-i\gamma^5\pi_a\tau_a\Big ]q
-\frac{1}{4G_1}\Big [\sigma_a\sigma_a+ \pi_a\pi_a\Big ]-
  \frac1{4G_2}\Big [\Delta^{s*}_{AA'}\Delta^s_{AA'}+
  \Delta^{p*}_{AA'}\Delta^p_{AA'}\Big ]
  \nonumber\\
&-&\frac{\Delta^{s*}_{AA'}}{2}[\bar
q^Ci\gamma^5\tau_A\lambda_{A'} q]
-\frac{\Delta^s_{AA'}}{2}[\bar q i\gamma^5\tau_A\lambda_{A'}
q^C]-\frac{\Delta^{p*}_{AA'}}{2}[\bar 
q^C\tau_A\lambda_{A'} q]
-\frac{\Delta^p_{AA'}}{2}[\bar q\tau_A\lambda_{A'} q^C],
\label{2}
\end{eqnarray}
where here and in the following the summation over  repeated
indices $a=0,...,8$ and $A,A'=2,5,7$ is implied.  Lagrangians
(\ref{1}) and (\ref{2}) are equivalent which simply follows from
the equations of motion for the bosonic fields
\begin{eqnarray}
&&\sigma_a(x)=-2G_1(\bar
q\tau_aq),~~~\Delta^s_{AA'}(x)=-2G_2(\bar
q^Ci\gamma^5\tau_A\lambda_{A'}q),~~~
\Delta^{s*}_{AA'}(x)=-2G_2(\bar qi\gamma^5\tau_A\lambda_{A'} q^C),
\nonumber\\&&\pi_a(x)=-2G_1(\bar
qi\gamma^5\tau_a q),~~~ \Delta^p_{AA'}(x)=-2G_2(\bar
q^C\tau_A\lambda_{A'}q),~~~
\Delta^{p*}_{AA'}(x)=-2G_2(\bar q \tau_A\lambda_{A'} q^C).
\label{3}
\end{eqnarray}
In (2)-(3)  $\sigma_a (x),\Delta^{s}_{AA'}(x)$ and $\pi_a
(x),\Delta^{p}_{AA'}(x)$ are scalar and pseudoscalar fields,
correspondingly.

Let us consider the flavor group SU(3)$_f$=SU(3)$_{L+R}$, which is
the diagonal subgroup of the chiral group. Then, all complex scalar
diquark fields
$\Delta^s_{AA'}(x)$ form an $(\bar 3_c,\bar 3_f)$ multiplet of the
SU(3)$_c\times$SU(3)$_f$ group, i.e. they are a color and flavor
antitriplet. The same is true for complex pseudoscalar diquark fields
$\Delta^p_{AA'}(x)$ which are also the components of an $(\bar
3_c,\bar 3_f)$-multiplet of the SU(3)$_c\times$SU(3)$_f$ group.
Evidently, all diquarks $\Delta^{s,p}_{AA'}(x)$ have the same
nonzero baryon charge. All the real $\sigma_a(x)$ and $\pi_a(x)$
fields are color singlets. Moreover, the set of scalar $\sigma
_a(x)$ mesons is decomposed into a direct sum of the singlet and
octet representations of the diagonal flavor group SU(3)$_f$. The
same decomposition into multiplets is true for the set of all
pseudoscalar $\pi_a(x)$ mesons. Clearly, in this case the octet is
constructed from three pions ($\pi^{\pm}$ and $\pi^0$), four kaons
($K^0$, $\bar K^0$ and $K^\pm$) and the eta-meson ($\eta_8$),
whereas the singlet ($\eta_0$) corresponds to the
$\eta^{\,\prime}$ meson.

In our previous paper \cite{ek}, using the intermediate 
bosonic Lagrangian (\ref{2}) and the Nambu--Gorkov formalism we
have obtained in the one-fermion loop approximation the effective
action ${\cal S}_{\rm {eff}}$ of the initial model (\ref{1}). In
terms of collective bosonic fields (\ref{3}) it takes the following
form:
\begin{equation}
{\cal S}_{\rm {eff}}(\sigma
_a,\pi_a,\Delta^{s,p}_{AA'},\Delta^{s,p*}_{AA'})
=-\int d^4x\left[\frac{\sigma^2_a+\pi^2_a}{4G_1}+
\frac{\Delta^s_{AA'}\Delta^{s*}_{AA'}+\Delta^p_{AA'}\Delta^{p*}_{AA'}
}{4G_2}\right]-\frac i2{\rm Tr}_{scfxNG}\ln Z, 
\label{4}
\end{equation}
where $Z$ is the $2\times 2$-matrix in the Nambu--Gorkov space,
\begin{equation}
Z=\left (\begin{array}{cc}
  D^+, & - K\\
- K^*, & D^-
\end{array}\right ),
\label{5}
\end{equation}
and the following notations are adopted
\begin{eqnarray}
&&D^+=i\gamma^\nu\partial_\nu+\mu\gamma^0-\Sigma,~~~~~~~
\Sigma=\tau_a\sigma_a+ i\gamma^5\pi_a\tau_a,~~~~~~~
K=(\Delta^p_{AA'}+i\Delta^s_{AA'}\gamma^5)\tau_A\lambda_{A'},
\nonumber\\&&
D^-=i\gamma^\nu\partial_\nu-\mu\gamma^0-\Sigma^t,~~~~~~
\Sigma^t=\tau_a^t\sigma_a+ i\gamma^5\pi_a\tau^t_a,~~~~~~
K^*=(\Delta^{p*}_{AA'}+i\Delta^{s*}_{AA'}\gamma^5)\tau_A
\lambda_{A'}.
\label{6}
\end{eqnarray}
Besides of an evident trace over the two-dimensional Nambu--%
Gorkov (NG) matrix, the Tr-operation in (\ref{4}) stands for 
the trace in spinor (s), flavor (f), color (c) as well as
four-dimensional coordinate (x) spaces, correspondingly. Let us
suppose that parity is
conserved so that all pseudoscalar diquark and meson fields have
zero ground state expectation values, i.e.
$\vev{\Delta^p_{AA'}(x)}=0$ and $\vev{\pi_a(x)}=0$. Furthermore,
since at zero $s$-quark mass, $m_s=0$, only the competition between
the chirally broken quark matter phase and the CFL one is relevant to
the physics of dense QCD (see, e.g., \cite{alford2}),
we permit in the present consideration nonzero ground state
expectation values only for $\sigma_0(x)$ and $\Delta^s_{AA}(x)$
fields ($A=2,5,7$). Namely, let $\vev{\sigma_0 (x)} \equiv\sigma,
~\vev{\Delta^{s}_{AA}(x)}\equiv\Delta$, $\vev{\Delta^{s*}_{AA}(x)}
\equiv\Delta^{*}$, where $A=2,5,7$, but other boson fields from
(\ref{3}) have zero ground state expectation values. In the case
$\Delta=0$, $\sigma\ne 0$ quark matter is in the chirally broken
phase, where the ground state is invariant under
SU(3)$_c\times$SU(3)$_f\times$U(1)$_B$. If $\Delta\ne 0$, then the
CFL phase is realized in the model, and the initial symmetry is
spontaneously broken down to SU(3)$_{L+R+c}$. 
\footnote{In spite of the fact that in the CFL phase the chiral
symmetry is also broken, the notation "chirally broken phase" is used
here and in the following for the phase without color
superconductivity.}
Now, let us make the following shifts of bosonic fields in (\ref{4}):
$\sigma_0 (x)\to\sigma_0(x)+\sigma$,
$\Delta^{s*}_{AA}(x)\to\Delta^{s*}_{AA} (x)+\Delta^*$,
$\Delta^{s}_{AA}(x) \to\Delta^{s}_{AA}(x)+\Delta$, 
($A=2,5,7$), and other bosonic fields remain unshifted. (Obviously,
the new shifted bosonic fields $\sigma_0 (x),\Delta^s_{AA}(x)$ etc,
now denote the (small) quantum fluctuations around the
mean values $\sigma,\Delta$ etc of mesons and diquarks rather
than the original fields (\ref{3}).) In this case 
\begin{equation}
Z\rightarrow\left (\begin{array}{cc}
D^+_o~, & -K_o\\
  -K_o^*~, & D^-_o
\end{array}\right )-\left (\begin{array}{cc}
\Sigma~, & K\\
  K^*~, & \Sigma^t
\end{array}\right )\equiv S_0^{-1}-\left (\begin{array}{cc}
\Sigma~, & K\\
  K^*~, & \Sigma^t
\end{array}\right ), 
\label{9}
\end{equation}
where $K_o,K^*_o, D^\pm_o,\Sigma_o,\Sigma^t_o$ are the corresponding
quantities (\ref{6}), in which all bosonic fields are replaced by
their own ground state expectation values, i.e. $\sigma_0(x)\to
\sigma$, $\pi_a(x)\to 0$,  $\Delta^{s}_{AA}(x)\to\Delta$,
$\Delta^{p}_{AA'}(x)\to 0$ etc, and $S_0$ is the quark propagator
matrix in the Nambu--Gorkov representation (its matrix elements
$S_{ij}$ are given in Appendix B). Then, expanding the obtained
expression into a Taylor-series up to second order of small bosonic
fluctuations, we have
\begin{equation}
{\cal S}_{\rm
{eff}}(\sigma_a,\pi_a,\Delta^{s,p}_{AA'},\Delta^{s,p*}_{AA'})=
{\cal S}_{\rm {eff}}^{(0)} +{\cal S}_{\rm  {eff}}^{(2)}
(\sigma_a,\pi_a,\Delta^{s,p}_{AA'},\Delta^{s,p*}_{AA'})+\cdots,
\label{10}
\end{equation}
where 
\begin{eqnarray}
  &&{\cal S}_{\rm {eff}}^{(0)}=-\int
  d^4x\left[\frac{\sigma\sigma}{4G_1}+
\frac{3|\Delta|^2}{4G_2}\right]+\frac i2{\rm
Tr}_{scfxNG}\ln \left (S_0\right )\equiv -\Omega(\sigma,\Delta,
  \Delta^{*})\int d^4x,   \label{11}\\
&&{\cal S}^{(2)}_{\rm {eff}}
(\sigma_a,\pi_a,\Delta^{s,p}_{AA'},\Delta^{s,p*}_{AA'})
  = -\int d^4x\left[\frac{\sigma^2_a+\pi^2_a}{4G_1}+
\frac{\Delta^s_{AA'}\Delta^{s*}_{AA'}+\Delta^p_{AA'}
\Delta^{p*}_{AA'}}{4G_2}\right]+\nonumber\\
&&~~~~~~~~~~~~~~~~~~~~~~+\frac i4{\rm
Tr}_{scfxNG}\left\{S_0\left (\begin{array}{cc}
\Sigma~, & K\\
  K^*~, & \Sigma^t
\end{array}\right )S_0\left (\begin{array}{cc}
\Sigma~, & K\\
  K^*~, & \Sigma^t
\end{array}\right )\right\},
   \label{12}
\end{eqnarray}
and $\Omega(\sigma,\Delta,\Delta^{*})$ is the thermodynamic potential
of the system. Notice that the term linear in meson and diquark
fields vanishes in (\ref{10}) due to the gap equations.

The detailed investigations of the thermodynamic potential, performed
in our previous paper \cite{ek} for the above accepted model
parameter set, shows that at $\mu<\mu_c\approx 330$
MeV the chirally broken quark matter phase with SU(3)$_c\times
$SU(3)$_{f}\times$U(1)$_B$ symmetric ground state is realized in the
model (in this case $\Delta=0$ and $M\approx 355$ MeV,  where
$M=\sqrt{2/3}~\sigma$ is the dynamical quark mass). However,  at
$\mu >\mu_c$ the CFL phase of dense baryonic matter arises. In this
phase $M=0$ and $\Delta$ varies with $\mu$ (see Fig. 2 in \cite{ek}).
Below we suppose that the gap $\Delta$ is a real nonnegative number.

In the following we will study the spectrum of meson/diquark
excitations both in the CFL and chirally broken phases of the NJL
model. Since particle masses are calculated by the use of
corresponding Green functions, it is necessary to put a special
attention to the effective action ${\cal S}_{\rm {eff}}^{(2)}$
(\ref{12}) which is really a generating functional of the
one-particle irreducible (1PI) two-point Green functions of mesons
and diquarks both in the chirally broken and CFL phases, namely 
\begin{eqnarray}
&& \Gamma_{XY}(x-y)=-\frac{\delta^2{\cal S}^{(2)}_{\rm eff}}{\delta
Y(y)\delta X(x)},
  \label{17}
\end{eqnarray}
where $X(x),Y(x)=\sigma_a(x),\pi_b(x),\Delta^{s,p}_{AA'}(x),
\Delta^{s,p*}_{BB'}(x)$. (To obtain a Green function (\ref{17}) in
the chirally broken phase of quark matter ($\mu<\mu_c$), one should
use in the expression for the quark propagator $S_0$ (see Appendix
B), entering in (\ref{12}), $M\approx 355$ MeV and $\Delta=0$,
whereas in the CFL phase ($\mu>\mu_c$) all Green functions (\ref{17})
correspond to $S_0$ with $M=0$ and values $\Delta\ne 0$ presented in
Fig. 2 of \cite{ek}.) In the following, we shall say that in the
theory there is a {\it mixing} between two different particles with
corresponding fields $X(x)$ and $Y(x)$, if their 1PI Green function
$\Gamma_{XY}(x-y)$ is not identically equal to zero. Now, after
performing in (\ref{12}) the trace operation over the
two-dimensional Nambu-Gorkov space, we obtain
\begin{eqnarray}
{\cal S}^{(2)}_{\rm {eff}}={\cal S}^{(2)}_{\rm mesons}+{\cal
S}^{(2)}_{\rm diquarks}+{\cal S}^{(2)}_{\rm mixed},
\label{7}
\end{eqnarray}
where
\begin{eqnarray}
\label{8}
  {\cal S}^{(2)}_{\rm mesons} \!\!\!\!&&\!\!=
  -\int d^4x\frac{\sigma_a^2+\pi^2_a}{4G_1}+
\frac i4{\rm Tr}_{scfx}
\left\{S_{11}\Sigma S_{11}\Sigma +2S_{12}\Sigma^tS_{21}\Sigma +
S_{22}\Sigma^t S_{22}\Sigma^t\right\},\\
\label{170}
  {\cal S}^{(2)}_{\rm diquarks} \!\!\!\!&&\!\!=
  -\int d^4x\frac{\Delta^s_{AA'}\Delta^{s*}_{AA'}+\Delta^p_{AA'}
\Delta^{p*}_{AA'}}{4G_2}
+\frac i4{\rm Tr}_{scfx}
\left\{S_{12}K^*S_{12}K^* +2S_{11}KS_{22}K^* +
S_{21}K S_{21}K\right\},\\
  {\cal S}^{(2)}_{\rm mixed} \!\!\!\!&&\!\!=
\frac i2{\rm Tr}_{scfx}
\left\{S_{11}\Sigma S_{12}K^* +S_{21}\Sigma S_{11}K +
S_{12}\Sigma^t S_{22}K^*+S_{21}KS_{22}\Sigma^t\right\},
  \label{171}
\end{eqnarray}
and $S_{ij}$ are the matrix elements of the quark propagator matrix
$S_0$ defined in (\ref{9}) (see also Appendix B).
(Some necessary explanations concerning the trace-operation over
coordinate space in the expressions (\ref{8})-(\ref{171}) are given
in Appendix A (see (\ref{B4}))). It follows from these formulae that
the effective action (\ref{8}) depends on the mesonic fields
$\sigma_a(x),\pi_b(x)$ only, i.e. it is a generating functional of
the two-point 1PI Green functions of mesons. Furthermore, the
effective action (\ref{170}) is composed from diquark fields only,
and the mixing between mesons and diquarks might occur because of the
effective action (\ref{171}). However, as a detailed analysis of the
NJL model (\ref{1}) with three massless quarks shows, each Green
function, containing mixing of mesons and diquarks, is proportional
to $M\Delta$. Thus, both in the chirally broken phase of quark
matter, where $\Delta=0$, and in the CFL one, where $M=0$, there is
no mixing between mesons and diquarks, and so the term (\ref{171}) 
can safely be ignored in our present consideration.
\footnote{Note, if some of the current quark masses are nonzero,
then due to $M\ne 0$ there arises a mixing between mesons and
diquarks in the CFL phase. This effect is analogous to the mixing
between the $\sigma$-meson and the scalar diquark in the color
superconducting phase of a two-flavor NJL model with nonzero masses
of $u$- and $d$-quarks \cite{eky1,eky2}.}

Finally, note that because of the traces containing an odd number of
$\gamma^5$ matrices, there is no mixing between scalar and
pseudoscalar particles in the framework of the NJL model (1),
as required by parity conservation.\label{II}

\section{Meson masses}

\subsection{The case of the CFL phase}

Taking into account the remarks from the previous section, we have
the following general expressions for the nonzero two-point 1PI
Green function of mesons which are valid in the chirally broken phase
of quark matter as well as in the CFL one (other two-point mesonic
1PI Green functions are zero in the model under consideration):
\begin{eqnarray}
&& \Gamma_{\sigma_a\sigma_a}(x-y)=-\frac{\delta^2{\cal S}^{(2)}_{\rm
mesons}}{\delta \sigma_a(y)\delta \sigma_a(x)},~~~~~~
\Gamma_{\pi_b\pi_b}(x-y)=-\frac{\delta^2{\cal S}^{(2)}_{\rm
mesons}}{\delta \pi_b(y)\delta \pi_b(x)}
  \label{172}
\end{eqnarray}
where $a,b=0,1,2,...,8$. In momentum space the zeros of the Fourier
transformations of these functions are connected with meson masses.

Starting from (\ref{172}), it is possible to obtain the expressions
for the mesonic 1PI Green functions ($a,b=0,1,2,...,8$)
\begin{eqnarray}
&& \Gamma_{\pi_a\pi_a}(z)=\frac{\delta(z)}{2G_1}+\frac
i2{\rm Tr}_{scf}
\left[S_{11}(z)\gamma^5\tau_aS_{11}(-z)\gamma^5\tau_a+S_{12}(z)
\gamma^5\tau^t_aS_{21}(-z)\gamma^5
\tau_a\right.\nonumber\\&&~~~~~~~~~~~~~~~~~~~~~~~~~~~~~~~\left.
+S_{21}(z)\gamma^5\tau_aS_{12}(-z)\gamma^5\tau^t_a+S_{22}(z)
\gamma^5\tau^t_aS_{22}(-z)\gamma^5\tau^t_a\right],\label{173}\\
&& \Gamma_{\sigma_b\sigma_b}(z)=\frac{\delta(z)}{2G_1}-\frac
i2{\rm Tr}_{scf}
\left[S_{11}(z)\tau_bS_{11}(-z)\tau_b+S_{12}(z)\tau^t_bS_{21}(-z)
\tau_b\right.\nonumber\\&&~~~~~~~~~~~~~~~~~~~~~~~~~~~~~~~\left.
+S_{21}(z)\tau_bS_{12}(-z)\tau^t_b+S_{22}(z)\tau^t_bS_{22}(-z)
\tau^t_b\right].
\label{174}
\end{eqnarray}
In (\ref{173})-(\ref{174}) $z=x-y$ and the matrix
elements $S_{ij}(z)$ are presented in formulae
(\ref{121})-(\ref{124}), from which the Fourier transformations
$\overline{S}_{ij}(p)$ are directly seen. The corresponding Fourier
transformations $\overline{\Gamma}_{\pi_a\pi_a}(p)$ and
$\overline{\Gamma}_{\sigma_b \sigma_b}(p)$ look like (as an example,
see the relation (\ref{B7}) from Appendix A):
\begin{eqnarray}
\label{175}
&&\overline{\Gamma}_{\pi_a\pi_a}(p)=\frac{1}{2G_1}+\frac i2{\rm
Tr}_{scf}\int\frac{d^4q}{(2\pi)^4}\left[~\overline{S}_{11}(p+q)
\gamma^5\tau_a\overline{S}_{11}(q)\gamma^5\tau_a+
\overline{S}_{12}(p+q)\gamma^5\tau^t_a
\overline{S}_{21}(q)\gamma^5\tau_a\right.\nonumber\\&&~~~~~~~~~~~~~~~
~~~~~~~~~~~~~~~~\left.+\overline{S}_{21}(p+q)\gamma^5\tau_a
\overline{S}_{12}(q)\gamma^5\tau^t_a+\overline{S}_{22}(p+q)\gamma^5
\tau^t_a\overline{S}_{22}(q)\gamma^5\tau^t_a\right],\\
&&\overline{\Gamma}_{\sigma_b\sigma_b}(p)=\frac{1}{2G_1}-\frac
i2{\rm
Tr}_{scf}\int\frac{d^4q}{(2\pi)^4}\left[~\overline{S}_{11}(p+q)\tau_b
\overline{S}_{11}(q)\tau_b+\overline{S}_{12}(p+q)\tau^t_b
\overline{S}_{21}(q)\tau_b\right.\nonumber\\&&~~~~~~~~~~~~~~~~~~~~~~~
~~~~~~~~\left.+\overline{S}_{21}(p+q)\tau_b\overline{S}_{12}(q)
\tau^t_b+\overline{S}_{22}(p+q)\tau^t_b\overline{S}_{22}(q)
\tau^t_b\right].
\label{176}
\end{eqnarray}
The zeros of these functions determine the $\pi$- and $\sigma$-meson
dispersion laws, i.e. the relations between their energy
and three-momenta. In the present paper, we are mainly interested
in the investigation of the modification of meson and diquark masses
in dense and cold quark matter. Since in this case a particle mass is
defined as the value of its energy in the rest frame, $\vec p=0$
(see, e.g., \cite{eky1,eky2,he,ruivo}), we put $p=(p_0,0,0,0)$ in the
following. As a result, the calculation of two-point 1PI Green
functions is significantly simplified. In order to perform for the
case of the CFL phase in (\ref{175})-(\ref{176}) the trace operations
over color and flavor spaces, we used the program of analytical
calculations MAPLE. The trace over spinor space and the subsequent
integration over $q_0$, has been performed by applying the
technique elaborated in \cite{eky1,eky2}. As a result, we have for
the 1PI meson Green functions with $a=1,2,...,8$:
\begin{eqnarray}
&& \overline{\Gamma}_{\sigma_a\sigma_a}(p_0)=\frac{1}{2G_1}+{\cal
A}-{\cal B},~~~~~~~~
\overline{\Gamma}_{\pi_a\pi_a}(p_0)=\frac{1}{2G_1}+{\cal A}+{\cal
B}, \label{4.1}
\end{eqnarray}
where
\begin{eqnarray}
&& {\cal A}=\int\frac{d^3q}{(2\pi)^3}
\left\{\frac{28(E_{\Delta}^++E_{\Delta}^-)[
E_{\Delta}^+E_{\Delta}^-+E^+E^-]}{3E_{\Delta}^+E_{\Delta}^-[p_0^2-
(E_{\Delta}^++E_{\Delta}^-)^2]}+\frac{4(E_{2\Delta}^++E_{\Delta}^-)[
E_{2\Delta}^+E_{\Delta}^-+E^+E^-]}{3E_{2\Delta}^+E_{\Delta}^-[p_0^2-
(E_{2\Delta}^++E_{\Delta}^-)^2]}+\right.\nonumber\\
&&~~~~~~~~~\left.+\frac{4(E_{\Delta}^++E_{2\Delta}^-)[
E_{\Delta}^+E_{2\Delta}^-+E^+E^-]}{3E_{\Delta}^+E_{2\Delta}^-[p_0^2-
(E_{\Delta}^++E_{2\Delta}^-)^2]}\right\},\label{4.4}\\
&&{\cal B}=\int\frac{d^3q}{(2\pi)^3}
\left\{\frac{8\Delta^2(E_{\Delta}^++
E_{\Delta}^-)}{3E_{\Delta}^+E_{\Delta}^-[p_0^2-
(E_{\Delta}^++E_{\Delta}^-)^2]}+\frac{8\Delta^2(E_{2\Delta}^++
E_{\Delta}^-)}{3E_{2\Delta}^+E_{\Delta}^-[p_0^2-
(E_{2\Delta}^++E_{\Delta}^-)^2]}+\right.\nonumber\\
&&~~~~~~~~~\left.+\frac{8\Delta^2(E_{\Delta}^++
E_{2\Delta}^-)}{3E_{\Delta}^+E_{2\Delta}^-[p_0^2-
(E_{\Delta}^++E_{2\Delta}^-)^2]}\right\}.
\label{4.5}
\end{eqnarray}
Moreover, we use in these formulae the notations
$(E_{\Delta}^\pm)^2=(E^\pm)^2+|\Delta|^2$, 
$(E_{2\Delta}^\pm)^2=(E^\pm)^2+4|\Delta|^2$, $E^\pm=E\pm\mu$,
$E=\sqrt{\strut\vec q^2+M^2}$, in which
$M$ is set equal to zero. It is clear from (\ref{4.4})-(\ref{4.5})
that each of the Green functions (\ref{4.1}) depends on $p_0^2$. So
the mass squared of the $a$-th scalar (or pseudoscalar) meson
($a=1,2,...,8$) is determined by a zero of the function
$\overline{\Gamma}_{\sigma_a\sigma_a}(p_0)$ (or
$\overline{\Gamma}_{\pi_a\pi_a}(p_0)$) in the  $p_0^2$ plane.
Moreover, it is evident from (\ref{4.1}) that all scalar mesons with
$a=1,2,...,8$ have the same mass in the CFL phase, thus forming an
SU(3) octet of scalar mesons. In a similar way, all pseudoscalar
mesons with $a=1,2,...,8$ form another massive SU(3) octet as well.
The masses of scalar and pseudoscalar meson octets in the CFL phase
are presented in Fig. 1 and 2, correspondingly.

In contrast, the two-point Green functions for $\sigma_0(x)$ and
$\pi_0(x)$ mesons take another form. Indeed,
\begin{eqnarray}
&& \overline{\Gamma}_{\sigma_0\sigma_0}(p_0)=\frac{1}{2G_1}+{\cal
Q}+{\cal R},~~~~~~~~
\overline{\Gamma}_{\pi_0\pi_0}(p_0)=\frac{1}{2G_1}+{\cal Q}-{\cal
R}, \label{4.6}
\end{eqnarray}
where
\begin{eqnarray}
&& {\cal
Q}=\int\frac{d^3q}{(2\pi)^3}\left\{\frac{32(E_{\Delta}^++E_{\Delta}^-
)[E_{\Delta}^+E_{\Delta}^-+E^+E^-]}{3E_{\Delta}^+E_{\Delta}^-[p_0^2-
(E_{\Delta}^++E_{\Delta}^-)^2]}+\frac{4(E_{2\Delta}^++E_{2\Delta}^-
)[E_{2\Delta}^+E_{2\Delta}^-+E^+E^-]}{3E_{2\Delta}^+E_{2\Delta}^-[p_0
^2-(E_{2\Delta}^++E_{2\Delta}^-)^2]}\right\},\label{4.7}\\
&&{\cal
R}=\int\frac{d^3q}{(2\pi)^3}\left\{\frac{32\Delta^2(E_{\Delta}^++
E_{\Delta}^-)}{3E_{\Delta}^+E_{\Delta}^-[p_0^2-
(E_{\Delta}^++E_{\Delta}^-)^2]}+\frac{16\Delta^2(E_{2\Delta}^++
E_{2\Delta}^-)}{3E_{2\Delta}^+E_{2\Delta}^-[p_0^2-
(E_{2\Delta}^++E_{2\Delta}^-)^2]}\right\}
\label{4.8}
\end{eqnarray}
(in (\ref{4.7})-(\ref{4.8}) the quantities $E^\pm$ etc are taken
again at $M=0$). Evidently, these mesons are singlets with respect to
the SU(3) group, and their masses are presented also in Figs 1,2. It
is clear from these figures that none of the mesons have a zero mass
in the CFL phase, i.e. they are not the Nambu -- Goldstone bosons
(NG) of this phase. Moreover, one can see that in the CFL phase there
is a singlet-octet mass splitting of pseudoscalar and scalar mesons,
which however vanishes in the $\Delta =0$ limit. Indeed, if the value
$\Delta =0$ is used in (\ref{4.1})-(\ref{4.8}), then the 1PI Green
functions of the octet and singlet mesons are the same, i.e. the mass
splitting is absent.
\begin{figure}
  \includegraphics[width=0.45\textwidth]{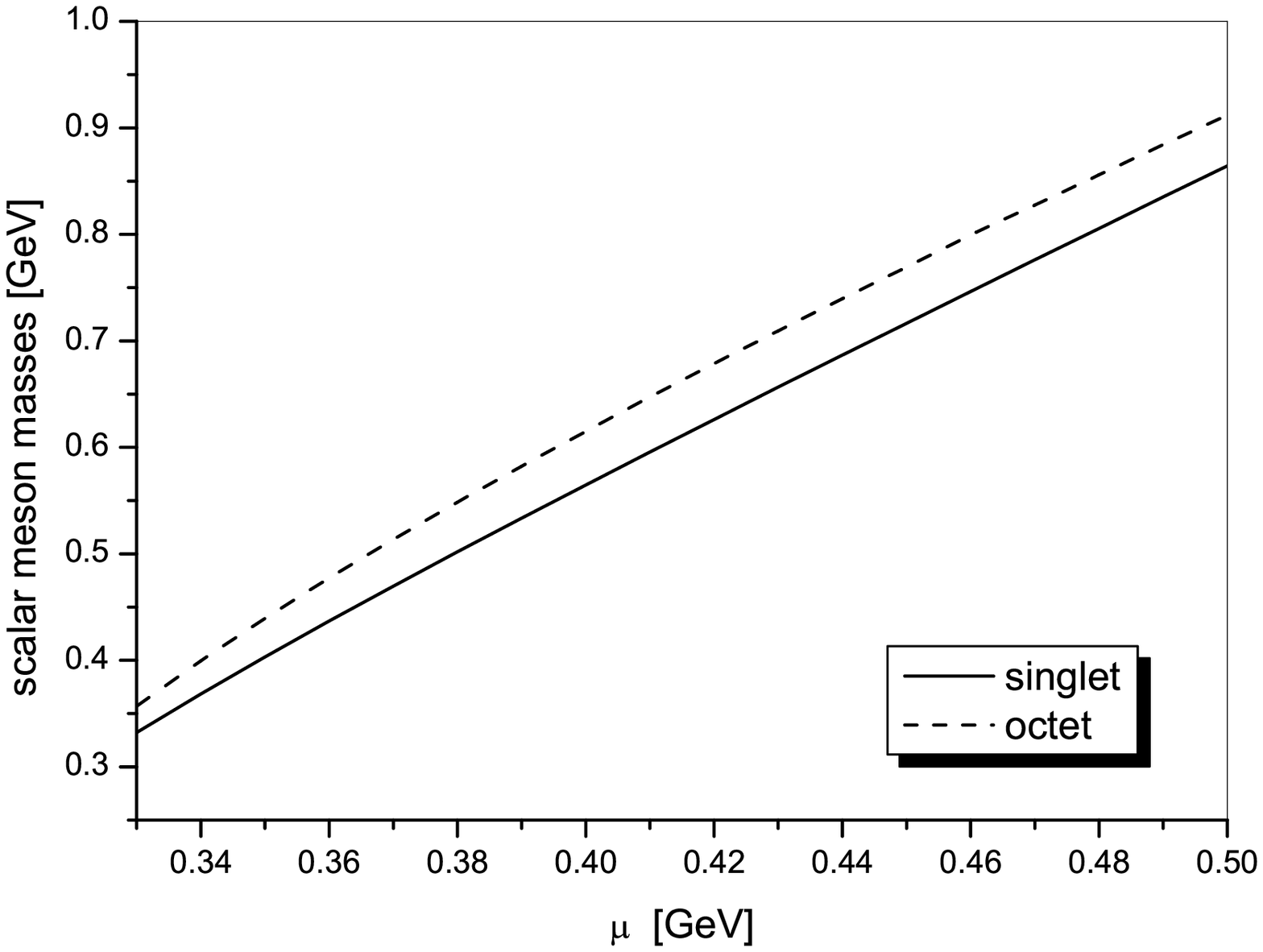}
  \hfill
  \includegraphics[width=0.45\textwidth]{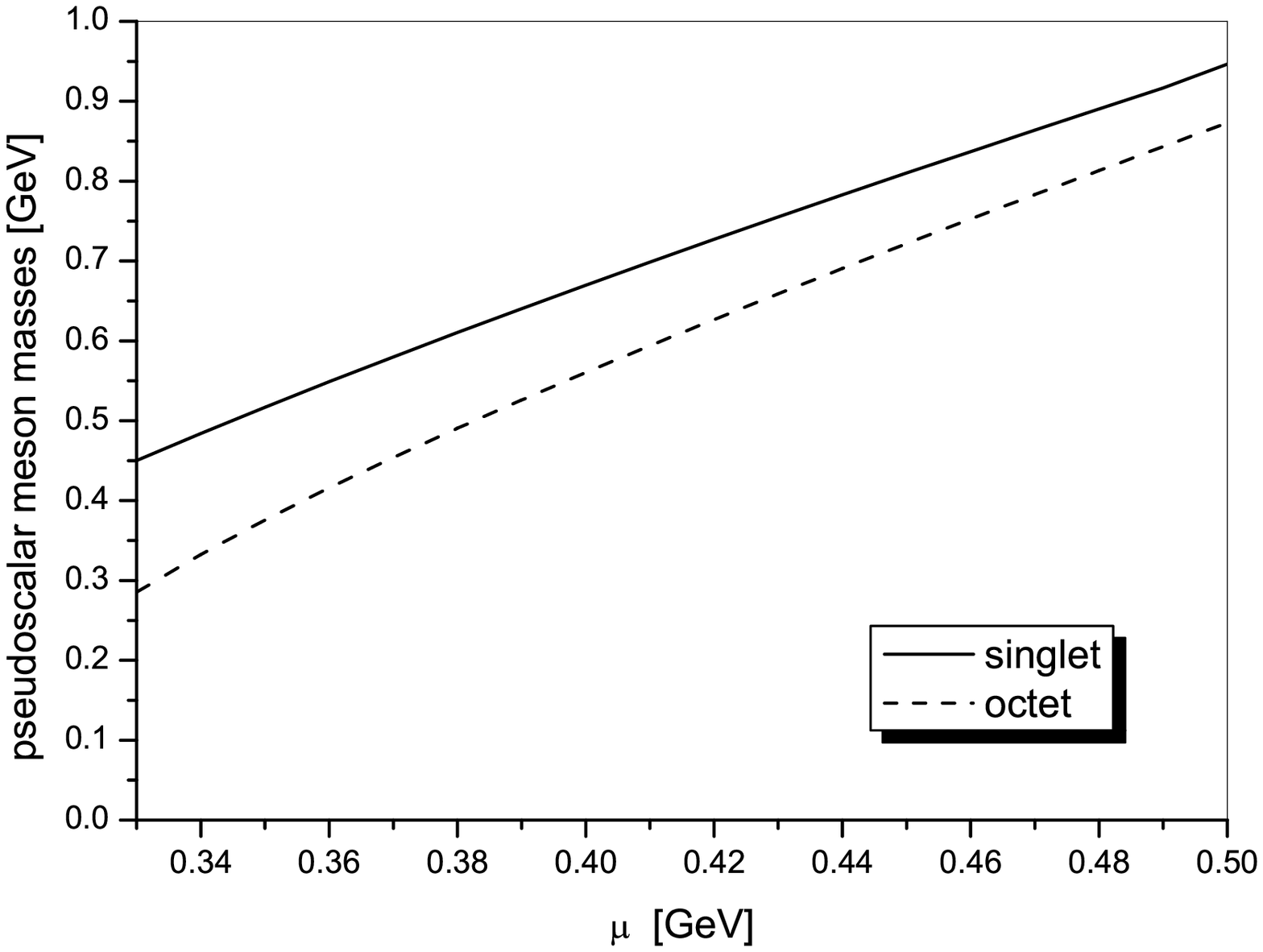}\\
\parbox[t]{0.45\textwidth}{
  \caption{The behaviour of the scalar meson masses vs $\mu$ in the
CFL phase.} \label{fig:1} }
\hfill
\parbox[t]{0.45\textwidth}{
\caption{The behaviour of the pseudoscalar meson masses vs $\mu$ in
  the CFL phase.}
\label{fig:2} }
\end{figure}

\subsection{The case of chirally broken quark matter phase}

Now, let us find the meson masses in the chirally broken phase of 
quark matter, i.e. at at $\mu<\mu_c$, $M\approx 355$ MeV, and $\Delta
=0$. In this case the calculation of the traces over
color and flavor indices in (\ref{175})-(\ref{176}) is greatly
simplified, so in the rest frame, $p=(p_0,0,0,0)$, the mesonic Green
functions look like ($a,b=0,1,2,...,8$): 
\begin{eqnarray}
&&\overline{\Gamma}_{\pi_a\pi_a}(p_0)=p_0^2\int\frac{d^3q}{(2\pi)^3}
\cdot\frac{12}{E[p_0^2-4E^2]},~~~~~~~\overline{\Gamma}_{\sigma_b
\sigma_b}(p_0)=(p_0^2-4M^2)\int\frac{d^3q}{(2\pi)^3}
\cdot\frac{12}{E[p_0^2-4E^2]}.
  \label{18}
\end{eqnarray}
(In obtaining (\ref{18}), the gap equation (see, e.g., \cite{ek}) was
used in order to eliminate the coupling constant $G_1$ from the
expressions (\ref{175})-(\ref{176}). 
Note also, the expressions (\ref{18}) do not follow directly from
(\ref{4.1})-(\ref{4.8}) in the $\Delta =0$ limit.) Evidently, in the
chirally broken phase the
Green functions $\overline{\Gamma}_{\pi_a\pi_a}(p_0)$ turn into zero
at $p_0^2=0$ for all $a=0,1,...,8$. It means that nine massless
excitations, Nambu --
Goldstone bosons, do exist in the pseudoscalar meson sector of the
model in the chirally broken phase. This fact corresponds to a
spontaneous symmetry breaking of the initial SU(3)$_L\times$SU(3)$
_R\times$SU(3)$_c\times$U(1)$_B\times$U(1)$_A$ down to the subgroup
SU(3)$_c\times$SU(3)$_{f}\times$U(1)$_B$ in the chirally broken
phase. Moreover, it is clear from (\ref{18}) that in this phase there
exists a nonet of scalar mesons with mass $\equiv 2M\sim 710$ MeV.

In summary, we have three main conclusions about mesons in the
framework of the NJL model (1) with three massless quarks. Firstly,
we see that nine pseudoscalar mesons are NG bosons only in the 
chirally broken phase of quark matter. In the CFL phase they cease
to be NG bosons, since now they acquire finite masses (see Fig. 2).
Secondly, the CFL breaking of the symmetry generates the
octet-singlet mass splitting of mesons. Thirdly, in the CFL phase the
mass splitting among scalar and pseudoscalar mesons occurs
differently. Indeed, at $\mu>\mu_c$ the mass of the scalar octet
mesons is larger than the mass of the scalar singlet meson (see Fig.
1), whereas for pseudoscalar mesons the situation is
inverse (see Fig. 2). \label{IIIB}

\section{Diquark masses}

\subsection{The case of the CFL phase}
\label{IVA}

As it follows from the discussion in section \ref{II}, all 
nonzero two-point 1PI Green function of diquarks both in the chirally
broken and CFL phases might be determined through the relation
\begin{eqnarray}
&& \Gamma_{XY}(x-y)=-\frac{\delta^2{\cal S}^{(2)}_{\rm
diquarks}}{\delta Y(y)\delta X(x)}, \label{180}
\end{eqnarray}
where the effective action ${\cal S}^{(2)}_{\rm diquarks}$ is given
in (\ref{170}),  and $X(x),Y(x)=\Delta^{s}_{AA'}(x),
\Delta^{s*}_{BB'}(x)$ or $X(x),Y(x)=\Delta^{p}_{AA'}(x),
\Delta^{p*}_{BB'}(x)$. The last restriction again means that scalar
and pseudoscalar diquarks do not mix in accord with parity
conservation.

As shown in our earlier paper \cite{eky2} for the case of the
two-flavor NJL model, any two-point Green function of pseudoscalar
diquarks differs in the color superconducting phase from a
corresponding Green function of scalar diquarks by a term which is
proportional to $M^2$, where $M$ is the dynamical quark mass in this
phase. The same is true for the CFL phase of the NJL model (1). 
So 
one can conclude that in the CFL phase of the model (1), where $M=0$,
each Green function of pseudoscalar diquarks is
equal to the corresponding Green function of scalar diquarks, e.g.,
$\Gamma_{\Delta^{p}_{AA'}\Delta^{p*}_{BB'}}(x-y)=
\Gamma_{\Delta^{s}_{AA'}\Delta^{s*}_{BB'}}(x-y)$, etc. 
Hence, to establish the spectrum of the diquark excitations of
the CFL phase, it is enough to study the set of scalar diquarks (the
mass spectrum of the pseudoscalar diquark excitations will be the
same in the CFL phase).

Let us consider the two-point 1PI Green functions of the scalar
diquarks. A more detailed analysis of the effective action
(\ref{170}) shows that in the CFL phase, i.e. at $\mu>\mu_c$, where
$M=0$ and $\Delta\ne 0$, eighteen scalar diquarks (nine
$\Delta^s_{AA'}(x)$ and nine $\Delta^{s*}_{AA'}(x)$ fields) may be
divided into four sectors: $s(57,75)$, $s(25,52)$, $s(27,72)$ and
$s(257)$. Each of the sectors $s(AA',A'A)$, where $A\ne A'$, is
composed of $\Delta^s_{AA'}(x)$, $\Delta^{s*}_{AA'}(x)$,
$\Delta^s_{A'A}(x)$, and $\Delta^{s*}_{A'A}(x)$ diquark fields, 
whereas the sector $s(257)$ is composed of six fields,
$\Delta^{s*}_{22}(x)$, $\Delta^{s*}_{55}(x)$, $\Delta^{s*}_{77}(x)$,
$\Delta^{s}_{22}(x)$, $\Delta^{s}_{55}(x)$, and $\Delta^{s}_{77}(x)$.
It turns out that there is a mixing between diquarks entering in the
same sector, whereas fields from different sectors are not mixed.
(The analogous situation takes place for the set of pseudoscalar
diquarks, which is divided into nonmixing sectors $p(57,75)$,
$p(25,52)$, $p(27,72)$ and $p(257)$.)

\subsubsection{The case of $s(AA',A'A)$ sectors}

Let us first study the mass spectrum of the excitations, e.g., in
the sector $s(57,75)$. The two-point 1PI Green functions of
scalar diquarks from this sector can be obtained from
(\ref{170}) by the relation (\ref{180}). In the rest frame, where
$p=(p_0,0,0,0)$, the Fourier transforms of these 1PI Green functions
form the following matrix \cite{ek}:
\begin{equation}
\overline{\Gamma}_{57,75}(p_0)=\mat {cccc}
\overline{\Gamma}_{\Delta^s_{57}\Delta^{s}_{57}}(p_0)&
\overline{\Gamma}_{\Delta^s_{57}\Delta^{s*}_{57}}(p_0)&
\overline{\Gamma}_{\Delta^s_{57}\Delta^{s}_{75}}(p_0)&
\overline{\Gamma}_{\Delta^s_{57}\Delta^{s*}_{75}}(p_0)\\
\overline{\Gamma}_{\Delta^{s*}_{57}\Delta^s_{57}}(p_0)&
\overline{\Gamma}_{\Delta^{s*}_{57}\Delta^{s*}_{57}}(p_0)&
\overline{\Gamma}_{\Delta^{s*}_{57}\Delta^s_{75}}(p_0)&
\overline{\Gamma}_{\Delta^{s*}_{57}\Delta^{s*}_{75}}(p_0)\\
\overline{\Gamma}_{\Delta^s_{75}\Delta^{s}_{57}}(p_0)&
\overline{\Gamma}_{\Delta^s_{75}\Delta^{s*}_{57}}(p_0)&
\overline{\Gamma}_{\Delta^s_{75}\Delta^{s}_{75}}(p_0)&
\overline{\Gamma}_{\Delta^s_{75}\Delta^{s*}_{75}}(p_0)\\
\overline{\Gamma}_{\Delta^{s*}_{75}\Delta^{s}_{57}}(p_0)&
\overline{\Gamma}_{\Delta^{s*}_{75}\Delta^{s*}_{57}}(p_0)&
\overline{\Gamma}_{\Delta^{s*}_{75}\Delta^{s}_{75}}(p_0)&
\overline{\Gamma}_{\Delta^{s*}_{75}\Delta^{s*}_{75}}(p_0)\emat
\equiv \mat {cccc}
0&A&C&0\\
B&0&0&C\\
C&0&0&A\\
0&C&B&0\emat,
\label{19}
\end{equation}
where $A\equiv\alpha+p_0\beta$, $B\equiv\alpha-p_0\beta$ and 
\begin{eqnarray}
&&\alpha=\int\frac{d^3q}{(2\pi)^3}
\left\{\frac{6E_{\Delta}^+p_0^2-(E_{\Delta}^++E_{2\Delta}^+)^2(
2E_{\Delta}^++E_{2\Delta}^+)}{9E_{\Delta}^+E_{2\Delta}^+[p_0^2-(
E_{\Delta}^++E_{2\Delta}^+)^2]}
+\frac{4p_0^2+4(E_{\Delta}^+)^2-10\Delta^2}
{3E_{\Delta}^+[p_0^2-4(E_{\Delta}^+)^2]}\right\}+\nonumber\\
&&~~~~~~~~~~~~~~~~~~~~~~~+\int\frac{d^3q}{(2\pi)^3}\bigg
\{E_{\Delta}^+\to
E_{\Delta}^-,~~E_{2\Delta}^+\to E_{2\Delta}^-\bigg\}, 
\label{21}\\
&&\beta=\int\frac{d^3q}{(2\pi)^3}
\left\{\frac{E^+(E_{\Delta}^++E_{2\Delta}^+)}{3E_{\Delta}^+E_{2\Delta
}^+[p_0^2-(E_{\Delta}^++E_{2\Delta}^+)^2]}+\frac{10E^+}{3E_{\Delta}^+
[p_0^2-4(E_{\Delta}^+)^2]}\right\}-\nonumber\\
&&~~~~~~~~~~~~~~~~~~~~~~-\int\frac{d^3q}{(2\pi)^3}\bigg
\{E^+\to E^-,~~E_{\Delta}^+\to
E_{\Delta}^-,~~E_{2\Delta}^+\to E_{2\Delta}^-\bigg\}, 
\label{22}\\
&&C=\int\frac{d^3q}{(2\pi)^3}
\left\{\frac{2\Delta^2(E_{\Delta}^++E_{2\Delta}^+)}{3E_{\Delta}^+
E_{2\Delta}^+[p_0^2-(E_{\Delta}^++E_{2\Delta}^+)^2]}-
\frac{10\Delta^2}{3E_{\Delta}^+
[p_0^2-4(E_{\Delta}^+)^2]}\right\}+\nonumber\\
&&~~~~~~~~~~~~~~~~~~~~~~+\int\frac{d^3q}{(2\pi)^3}\bigg\{E_{\Delta}^+
\to E_{\Delta}^-,~~E_{2\Delta}^+\to E_{2\Delta}^-\bigg\}.
\label{23}
\end{eqnarray}
(To obtain the above expressions for $\alpha$ and $\beta$, one has to
use the gap equation for $\Delta$ (see in \cite{ek}) in order to
eliminate the coupling constant $G_2$ from corresponding 1PI Green
functions.) Evidently, in the case of mixing between particles the
information about their masses should be extracted from the zeros of
the determinant of the matrix, composed from corresponding 1PI Green
functions. So, in our case it is necessary to study the equation
${\rm det}\overline{\Gamma}_{57,75}(p_0)=0$, which takes the
following form
\begin{equation}
{\rm det}\overline{\Gamma}_{57,75}(p_0)\equiv
(AB-C^2)^2=[(\alpha-C)(\alpha+C)-p_0^2\beta^2]^2=0.
\label{24}
\end{equation}
In the $p_0^2$-plane, each zero of this equation defines a mass
squared of a bosonic excitation of the CFL phase ground state in the
$s(57,75)$ sector. Since this sector contains four scalar
diquarks, one should search for four solutions of the equation
(\ref{24}) in the $p_0^2$-plane. Clearly, due to the structure of
(\ref{24}), this equation admits at least two different solutions
(each being two-fold degenerate), which are given by the zeros of the
expression in the square bracket. It was proved in \cite{ek} that
$\alpha-C\sim p_0^2$. Hence, the square bracket in (\ref{24}) becomes
zero at the point $p_0^2=0$. So, in the $s(57,75)$ sector there are
two massless scalar excitations, i.e. NG bosons.

Note, the expression ${\rm det}\overline{\Gamma}_{57,75}$ which is in
the left hand side of the equation (\ref{24}) is a
complex-valued function defined on the Riemann
manifold composed of an infinitely large number of sheets of the
variable $p_0^2$. The first (physical) sheet is the $p_0^2$ plane 
with the cut $4\Delta^2 <p_0^2$ along the real axis. (Just the
integrals (\ref{21})-(\ref{23}) supply us with the values of the
function ${\rm det}\overline{\Gamma}_{57,75}$ on this physical
sheet.) It turns out that apart from the trivial zero, $p_0^2=0$,
there are no solutions of the equation (\ref{24}) on this
sheet, so there are no stable massive diquark excitations in the
$s(57,75)$ sector. Using the procedure of analytical continuation
presented in \cite{eky1} we could find for each value of the chemical
potential $\mu$ a complex point, lying on the second sheet of
$p_0^2$, where the function ${\rm det}\overline{\Gamma}_{57,75}$
turns into zero. Evidently, the
real and imaginary parts of this point correspond to the mass and
width of a resonance. Hence, as was pointed out above, in the
$s(57,75)$ sector there appears a twicely degenerated excitation of
the CFL phase, whose mass and width are presented in Fig. 3 as
functions of $\mu$.

A similar situation occurs in the other four-component sectors
$s(25,52)$ and $s(27,72)$. Namely, for both sectors the 1PI Green
function matrix has the form (\ref{19}). Hence, in each of these
sectors there are two NG bosons as well as two resonances with the
same mass and width, given in Fig. 3. \label{IVA1}
\begin{figure}
\includegraphics[width=0.45\textwidth]{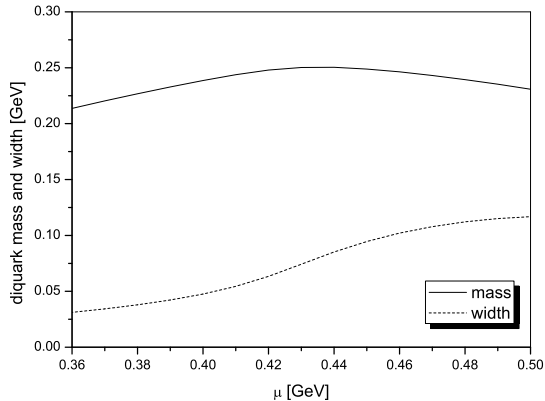}
\caption{The behaviour of the mass and the width of the scalar and
pseudoscalar diquark octets vs $\mu$ in the CFL phase.} \label{fig:3}
 \end{figure}

\subsubsection{Other diquark excitations of the CFL phase}

First of all note that there are 36 (18 scalar- and 18
pseudoscalar-) diquark fields (\ref{3}) in our model. So, there
should exist at least 36 elementary diquark excitations both in the
chirally broken quark matter phase and in the CFL one. The masses of
12 (six of them are NG bosons, the other six are massive resonances)
scalar diquark excitations of the CFL phase were obtained in the
previous section.

As it is clear from the discussion made at the beginning of section
\ref{IVA}, the mass spectrum of another 12 particles, which are 
the CFL ground state excitations in the pseudoscalar diquark sectors
$p(57,75)$, $p(25,52)$, and $p(27,72)$, is identical to the mass
spectrum of the corresponding scalar diquarks from sectors
$s(AA',A'A)$, where  $A\ne A'$ ( see the previous section
\ref{IVA1}). Hence, in addition to the scalar excitations found in
the previous section, in the CFL phase there exist six pseudoscalar
NG bosons as well as six pseudoscalar resonances, whose mass and
width are presented in Fig. 3. 

Concerning the CFL ground state diquark excitations from the sectors
$s(257)$ and $p(257)$, we should note that the corresponding 1PI
Green functions form in the rest frame of the momentum representation
a nontrivial 6$\times$6 matrix $\overline{\Gamma}_{257}(p_0)$, which
is the same both for the $s(257)$ and $p(257)$ diquark sectors, with
a rather complicated determinant, 
\begin{equation}
{\rm det}\overline{\Gamma}_{257}(p_0)=\big
[(P-W)(Q-Z)-(R-T)^2\big]^2\big\{(2T+R)^2-(2W+P)(2Z+Q)\big\}.
\label{290}
\end{equation}
An exact expression for the matrix
$\overline{\Gamma}_{257}(p_0)$ as well as for the 1PI Green functions
$P,Q,R,...$ from (\ref{290}) are presented in our previous paper
\cite{ek}, where it was also shown that the equation ${\rm
det}\overline{\Gamma}_{257} (p_0)=0$ has a three-fold degenerated
solution $p_0^2=0$. So in the diquark sectors $s(257)$ and $p(257)$
there are six (three scalar and three pseudoscalar) NG excitations,
and the initial NJL model (1) as a whole has 18 NG bosons in the mass
spectrum of the CFL phase.

A more detailed consideration of the quantities $P,Q,R,...$
shows that $P-W=A$, $Q-Z=B$, and $R-T=C$, where $A,B,C$ are given in
(\ref{19}). So, the square brackets in (\ref{290}) are no more than
${\rm det}\overline{\Gamma}_{57,75} (p_0)$ presented in (\ref{24}). 
\footnote{This fact was not observed in \cite{ek}, leading to an
incorrect statement about the multiplet structure of the massive
diquark excitations of the CFL phase.} 
As a consequence, we see, e.g., that in the
$s(257)$ sector of the model there exist two massive resonances with
the same mass and width, depicted in Fig. 3. Moreover, their mass and
width are identical to those for
the massive resonances from all scalar diquark sectors $s(AA',A'A)$.
So, all these scalar diquark resonances form in total an octet with
respect to the SU(3)$_{L+R+c}$ group. A similar situation is valid
for the pseudoscalar diquarks, where in the mass spectrum there is an
octet of the CFL phase excitations with the same mass and width (see
Fig. 3).

Unfortunately, we did not manage to find nontrivial diquark
excitations of the CFL phase from the sectors $s(257)$ and $p(257)$,
corresponding to a zero of the expression 
$F(p_0^2)\equiv (2T+R)^2-(2W+P)(2Z+Q)$ that appears in the braces of
(\ref{290}). (Evidently, both excitations are SU(3)$_{L+R+c}$
singlets.) The matter is that $F(p_0^2)$ is an analytical function on
a rather complicated Riemann manifold of the variable $p_0^2$. On its
first Riemann sheet there is only a trivial zero, $p_0^2=0$ (which
corresponds to the NG bosons, as it was discussed above). Due to a
rather complicated structure of the function $F(p_0^2)$, we were
not able to perform its continuation onto the second Riemann sheet
and get any information about the mass and width of the remaining
SU(3)$_{L+R+c}$ singlet diquark resonances from the sectors $s(257)$
and $p(257)$.

\subsection{The case of the chirally broken phase of quark matter}

In this phase, i.e. at $\mu<\mu_c$, the gap $\Delta$ vanishes, 
so the matrix elements $S_{12}(x-y)$ and $S_{21}(x-y)$
(see (\ref{122}) and (\ref{123}), correspondingly) of the quark
propagator matrix $S_0$ are vanishing, too. As a consequence, the
expression ${\cal S}^{(2)}_{\rm diquarks}$ of the two-point 1PI Green
functions for diquark fields is simplified in the chirally broken
phase:
\begin{eqnarray}
\label{0170}
  {\cal S}^{(2)}_{\rm diquarks} \!\!\!\!&&\!\!=
  -\int d^4x\frac{\Delta^s_{AA'}\Delta^{s*}_{AA'}+\Delta^p_{AA'}
\Delta^{p*}_{AA'}}{4G_2}
+\frac i2{\rm Tr}_{scfx}
\left\{S_{11}KS_{22}K^* \right\}.
\end{eqnarray}
Using this expression in (\ref{180}), it is possible to get for each
fixed $A,A'=2,5,7$:
\begin{eqnarray}
\label{0171}
  \Gamma_{\Delta^{s*}_{AA'}\Delta^{s}_{AA'}}(x-y) \!\!\!\!&&\!\!=
  \frac{\delta(x-y)}{4G_2}+\frac i2{\rm Tr}_{scf}
\left\{S_{11}(x-y)\gamma^5\tau_A\lambda_{A'}S_{22}(y-x)
\gamma^5\tau_A\lambda_{A'} \right\}\\
\label{0172}
  \Gamma_{\Delta^{p*}_{AA'}\Delta^{p}_{AA'}}(x-y) \!\!\!\!&&\!\!=
  \frac{\delta(x-y)}{4G_2}-\frac i2{\rm Tr}_{scf}
\left\{S_{11}(x-y)\tau_A\lambda_{A'}S_{22}(y-x)
\tau_A\lambda_{A'} \right\}.
\end{eqnarray}
In addition, the following relations are valid:
\begin{eqnarray}
\label{0173}
  \Gamma_{\Delta^{s}_{AA'}\Delta^{s*}_{AA'}}(x-y)=
  \Gamma_{\Delta^{s*}_{AA'}\Delta^{s}_{AA'}}(y-x),~~~~~
  \Gamma_{\Delta^{p}_{AA'}\Delta^{p*}_{AA'}}(x-y)=
  \Gamma_{\Delta^{p*}_{AA'}\Delta^{p}_{AA'}}(y-x),
\end{eqnarray}
and other two-point diquark 1PI Green functions are identically
equal to zero in the chirally broken phase. Using the expressions
(\ref{1210})-(\ref{1240}) for the fermion Green functions,
one can easily perform the Tr-operation over color and flavor indices
in (\ref{0171})-(\ref{0172}). Then, in the rest frame of the momentum
representation, i.e. at $p=(p_0,0,0,0)$, we have for each fixed pair
of $A,A'=2,5,7$:
\begin{eqnarray}
&&\overline{\Gamma}_{\Delta^{s*}_{AA'}\Delta^{s}_{AA'}}(p_0)=
\frac 1{4G_2}-16
\int\frac{d^3q}{(2\pi)^3}\frac{E}{4E^2-(p_0+2\mu)^2}\equiv
\frac 1{4G_2}-\Phi_s(\epsilon),\label{48}\\
&&\overline{\Gamma}_{\Delta^{p*}_{AA'}\Delta^{p}_{AA'}}(p_0)=
\frac 1{4G_2}-16\int\frac{d^3q}{(2\pi)^3}\frac{\vec
q^2}{E}\frac{1}{4E^2-(p_0+2\mu)^2}\equiv
\frac 1{4G_2}-\Phi_p(\epsilon),\label{49}
\end{eqnarray}
where $\epsilon=(p_0+2\mu)^2$. Moreover, it follows from
(\ref{0173}) that
$\overline{\Gamma}_{\Delta^{s}_{AA'}\Delta^{s*}_{AA'}}(p_0)$ $=$
$\overline{\Gamma}_{\Delta^{s*}_{AA'}\Delta^{s}_{AA'}}(-p_0)$,
$\overline{\Gamma}_{\Delta^{p}_{AA'}\Delta^{p*}_{AA'}}(p_0)$ $=$
$\overline{\Gamma}_{\Delta^{p*}_{AA'}\Delta^{p}_{AA'}}(-p_0)$.
From the above general consideration of the diquark Green
functions in the chirally broken quark matter phase we see that: i)
scalar diquarks do not mix with pseudoscalar ones, ii) each scalar
$\Delta^{s}_{AA'}$ or pseudoscalar $\Delta^{p}_{AA'}$ diquark field
is mixed only with its complex conjugated one. Hence, for each pair
of scalar diquarks $\Delta^{s}_{AA'},\Delta^{s*}_{AA'}$ (or
pseudoscalar diquarks $\Delta^{p}_{AA'},\Delta^{p*}_{AA'}$,
correspondingly) we have a simple 2$\times$2 matrix
$\overline{\Gamma}^s_{AA'}(p_0)$ of their 1PI Green functions (it is
a 2$\times$2 matrix $\overline{\Gamma}^p_{AA'}(p_0)$ for the system
of two pseudoscalar diquarks $\Delta^{p}_{AA'},\Delta^{p*}_{AA'}$,
correspondingly):
\begin{equation}
\overline{\Gamma}^{s}_{AA'}(p_0)= \left(\begin{array}[c]{cc}
0&\overline{\Gamma}_{\Delta^{s}_{AA'}\Delta^{s*}_{AA'}}(p_0)
\\
\overline{\Gamma}_{\Delta^{s*}_{AA'}\Delta^{s}_{AA'}}(p_0)&
0        \end{array}\right),~~~\overline{\Gamma}^{p}_{AA'}(p_0)=
\left(\begin{array}[c]{cc}
0&\overline{\Gamma}_{\Delta^{p}_{AA'}\Delta^{p*}_{AA'}}(p_0)
\\
\overline{\Gamma}_{\Delta^{p*}_{AA'}\Delta^{p}_{AA'}}(p_0)&
0        \end{array}\right).
\label{47}
\end{equation}
Actually, due to the relations (\ref{48})-(\ref{49}), these matrices
do not depend on $A,A'=2,5,7$, i.e. they are the same for each pair
of scalar $\Delta^{s}_{AA'},\Delta^{s*}_{AA'}$ or pseudoscalar
$\Delta^{p}_{AA'},\Delta^{p*}_{AA'}$ diquarks. Obviously, to obtain
the diquark excitations of the chirally broken phase, it is
sufficient to solve the equations ${\rm
det}\overline{\Gamma}^s_{AA'}(p_0)=0$ and
${\rm det}\overline{\Gamma}^p_{AA'}(p_0)=0$ or the following ones
\begin{eqnarray}
&&\overline{\Gamma}_{\Delta^{s*}_{AA'}\Delta^{s}_{AA'}}(p_0)\equiv
\frac 1{4G_2}-\Phi_s(\epsilon)=0,\label{50}\\
&&\overline{\Gamma}_{\Delta^{p*}_{AA'}\Delta^{p}_{AA'}}(p_0)
\equiv \frac 1{4G_2}-\Phi_p(\epsilon)=0.
\label{51}
\end{eqnarray}
In the present consideration we restrict
ourselves to looking only for stable diquark excitations of the
chirally broken phase.

Note, the functions $\Phi_s(\epsilon)$ and $\Phi_p(\epsilon)$ are
analytical in the whole complex $\epsilon$-plane, except for the cut
$4M^2<\epsilon$ along the real axis. (In general, these functions
are defined on complex Riemann surfaces which are to
be described by several sheets. The integral representations for
$\Phi_{s,p}(\epsilon)$, given in (\ref{48})-(\ref{49}), define its
values on the first sheet only. To find  values of
$\Phi_{s,p}(\epsilon)$ on the rest of the Riemann surfaces, a special
procedure of analytical continuation is needed (see, e.g., in
\cite{eky1}).) Let us denote by $\epsilon^s_0$ and $\epsilon^p_0$ the
solutions of the equations (\ref{50}) and (\ref{51}), respectively.
Of course, they depend on the coupling constant $G_2$ of the diquark
channel. Obviously, the stable diquark excitation corresponds to the
root $\epsilon^s_0$ which lies on the first Riemann sheet and obeys
the constraint $0<\epsilon^s_0<4M^2$. It is fulfilled only if
$H^*<G_2<H^{**}$, where $H^*$ and $H^{**}$ are defined by
\begin{eqnarray}
H^* &\equiv&\frac {1}{4\Phi_s(4M^2)}= \frac {\pi^2}{4\left
[\Lambda\sqrt{M^2+\Lambda^2}+M^2\ln((
\Lambda+\sqrt{M^2+\Lambda^2})/M)\right ]},\nonumber\\
H^{**}&\equiv&\frac {1}{4\Phi_s(0)}=\frac {\pi^2}{4\left
[\Lambda\sqrt{M^2+\Lambda^2}-M^2\ln((
\Lambda+\sqrt{M^2+\Lambda^2})/M)\right ]}=\frac{3G_1}{2}.
\label{52}
\end{eqnarray}
(Actually, the last equality in (\ref{52}), i.e. $H^{**}=3G_1/2$, is
due to the gap equation for $M\ne 0$.)
In this case $\epsilon^s_0$ is the mass squared of the stable scalar
diquark in the vacuum, i.e. at $\mu=0$. For a rather weak interaction
in the diquark channel ($G_2<H^*$),  $\epsilon^s_0$ runs onto the
second Riemann sheet, and unstable scalar diquark modes (resonances)
appear. Unlike this, a sufficiently strong interaction in the diquark
channel ($H^{**}<G_2$) pushes $\epsilon^s_0$ towards the negative
semi-axis of the first Riemann sheet, i.e. in this case
$\epsilon^s_0\equiv (M_D^o)^2<0$, where $M_D^o$ is the mass of the
diquark in the vacuum. The latter indicates a tachyon
singularity in the scalar diquark propagator, evidencing that the
SU(3)$_{L+R}\times$SU(3)$_c\times $U(1)$_B$ symmetric ground state of
the chirally broken phase is not stable (in this case there arises a
deeper ground state, corresponding to another phase of the model, the
CFL phase). A similar observation was made in the framework of a 
two-flavor NJL model, where the chirally broken quark matter phase is
unstable if there is a sufficiently strong interaction in the diquark
channel \cite{eky1,eky2,he2}. Indeed, at a very large  $G_2$, as
it has been shown in \cite{klim}, the color symmetry is spontaneously
broken even at a vanishing chemical potential.

Let us ignore for a moment the scalar diquark sector and perform a
similar analysis, based on the equation (\ref{51}), for pseudoscalar
diquark excitations. Then, pseudoscalar diquarks are stable
excitations of the chirally broken phase only, if the constraint
$H^{**}<G_2<H^{***}$ is fulfilled, where $H^{**}$ is given in
(\ref{52}) and
\begin{eqnarray}
&&H^{***}=\frac {1}{4\Phi_p(0)}=\frac
{\pi^2\Lambda\sqrt{M^2+\Lambda^2}}{4\left
[3M^2\Lambda^2+\Lambda^4-3M^2\Lambda\sqrt{M^2+\Lambda^2}\ln((
\Lambda+\sqrt{M^2+\Lambda^2})/M)\right ]}.
\label{gamma3p}
\end{eqnarray}
(In this case the solution $\epsilon^p_0$ of the
equation (\ref{51}) lies inside the interval $0<\epsilon^p_0<4M^2$.)
It turns out that at $G_2<H^{**}$ these excitations are unstable,
whereas at $H^{***}<G_2$ a tachyonic instability of the chirally
broken phase appears.

Now, combining together the above separate considerations of the
scalar and pseudoscalar diquark excitations, we may conclude that at
a rather weak interaction in the diquark channel ($G_2<H^{*}$) both
scalar and pseudoscalar diquark excitations of the chirally broken
phase are resonances. If $H^*<G_2<H^{**}$, then, in addition to
mesons, the scalar diquarks are stable particles in this phase (the
pseudoscalar diquarks are unstable as before). Note that the initial
massless NJL model (1) is parametrized by three independent
parameters $\Lambda$, $G_1$ and $G_2$. So, one may expect that
estimates $H^{*}$ and $H^{**}$ from (\ref{52}) depend on $\Lambda$
and $G_1$. However, as it was pointed out just after (\ref{52}), the
quantity $H^{**}\equiv 1.5 G_1$ does not depend really on the cutoff
parameter $\Lambda$. In contrast, $H^{*}$ depends both on $\Lambda$
and $G_1$. In particular, since for the parameter set accepted in
sec. II we
have $M\approx 0.355$ GeV, one can present in this case
the quantity $H^{*}$ in the following form $H^{*}\approx 0.660 G_1$.

Having a root $\epsilon^s_0\equiv (M_D^o)^2$ of the equation
(\ref{50}), one can find in the case $H^*<G_2<H^{**}$ two zeros
(with respect to the variable $p_0$) of the 1PI Green function
$\overline{\Gamma}_{\Delta^{s*}_{AA'}\Delta^{s}_{AA'}}(p_0)$ 
as well as four zeros of the equation ${\rm
det}\overline{\Gamma}^s_{AA'}(p_0)=0$. They provide us with the
following two different mass squared of the excitations in each two
scalar $\Delta^{s}_{AA'},\Delta^{s*}_{AA'}$-diquark system:
\begin{equation}
   \label{53}
(M_{\Delta})^2=(M_D^o-2\mu)^2,~~~~~(M_{\Delta^{*}})^2=
(M_D^o+2\mu)^2.
\end{equation}
In particular, for our choice of the model parameters (see sec. II)
we have $M_D^o\approx 1.968M$, where $M\approx 0.355$ GeV.
Furthermore, if 
$G_2\to H^{*}_+$ then $M_D^o\to 2M$, if $G_2\to H^{**}_-$ then
$M_D^o\to 0$. Since there are nine scalar $\Delta^{s}_{AA'}$ diquarks
as well as nine scalar $\Delta^{s*}_{AA'}$ antidiquarks in our model,
we relate $M_{\Delta}$ in (\ref{53}) to the mass of the diquark nonet
with the baryon number $B=2/3$ and $M_{\Delta^{*}}$ to the mass of
the antidiquark nonet with $B=-2/3$. The difference between diquark
and antidiquark masses in (\ref{53}) is explained by the absence of a
charge conjugation symmetry in the presence of a chemical potential
$\mu$.

Finally, if $H^{**}<G_2$, then a SU(3)$_{L+R}\times$SU(3)$_c\times
$U(1)$_B$ symmetric ground state, i.e. the chirally broken phase of
quark matter, is not allowed to exist in the model. The matter is
that in this case tachyon singularities of the scalar diquark
propagator alone (at $H^{**}<G_2<H^{***}$), or both of the scalar and
pseudoscalar diquark propagators (at $H^{***}<G_2$) appear. As a
result, in this case the ground state of the CFL phase is always
deeper in comparision with the ground state of the chirally broken
phase. So, only the CFL phase may be realized in the model at
sufficiently high values of the coupling constant $G_2$ and arbitrary
values of $\mu$. As a consequence, one must expect that at $G_2\to
H^{**}_-$ the critical value $\mu_c$ of the chemical potential tends
to zero. The fact that at $G_2\to H^{**}_-$ the diquark mass $M_D^o$
tends to zero may be considered as a precursor, which appears in the
chirally broken phase, of the spontaneous breaking of the SU(3)$_c$
symmetry, taking part at $G_2=H^{**}$.

\section{Summary and discussion}

In the present paper we have continued the investigation, started in
our previous paper \cite{ek}, of the bosonic excitations (mesons and
diquarks) of the dense quark matter, composed of $u$, $d$, and $s$
quarks, at zero temperature. The consideration is performed in the
framework of the massless NJL model (1), omitting the `t Hooft
six-quark interaction term, for simplicity. In this case, the initial
symmetry group of the model, i.e. SU(3)$_L\times$SU(3)$
_R\times$SU(3)$_c\times$U(1)$_B\times$U(1)$_A$ does contain the 
axial U(1)$_A$ subgroup. As a result, we have shown for the model
parameter set accepted in sec. II that at
sufficiently low values of $\mu$, $\mu<\mu_c\approx 330$ MeV, the
chirally broken quark matter phase with
SU(3)$_{L+R}\times$SU(3)$_c\times $U(1)$_B$-ground state symmetry is
realized and nine massless pseudoscalar mesons (which
are the NG bosons), $\pi^{\pm}$, $\pi^0$, $K^0$, $\bar K^0$,
$K^\pm$, $\eta_8$ and $\eta^{\,\prime}$, appear. (In massless QCD,
where U(1)$_A$ is broken on the quantum level, or in NJL models with
`t Hooft interaction the $\eta^{\,\prime}$-meson is not a NG boson.)

At $\mu>\mu_c$ the original symmetry is spontaneously
broken down to SU(3)$_{L+R+c}$, and the CFL phase does occur. In
this case, in accordance with the Goldstone theorem, eighteen
NG bosons must appear in the mass spectrum of the model (1). (In
contrast, due to the absence of the unphysical U(1)$_A$ symmetry,
only seventeen NG bosons must appear  in massless QCD.)
Considering 1PI Green functions, we have found nine NG
bosons in the sector of scalar diquark excitations. Eight of them
have to be considered as non-physical, since in real QCD they supply
masses to gluons by the Anderson -- Higgs mechanism. The remaining
scalar NG boson corresponds to a spontaneous breaking of the baryon
U(1)$_B$ symmetry. The other nine NG bosons are no more pseudoscalar
mesons, but now the massless excitations in the pseudoscalar diquark
sector of the model. All that, i.e. the NG boson structure of the
model (1), is the main result of the paper \cite{ek} which is also
confirmed in the present consideration. 

Besides NG diquarks, we have proved in \cite{ek} the existence of
massive diquark excitations in the CFL phase. However, a detailed
numerical analysis of the diquark masses vs chemical potential $\mu$
was not done there. In the present paper we have argued that all
massive diquark excitations of the CFL phase are resonances, since
the corresponding singularities of their Green functions in momentum
space lie on the second energy Riemann sheet. Moreover, they form
scalar- and pseudoscalar SU(3)$_{L+R+c}$ octets and singlets.
The mass and width of scalar- and pseudoscalar diquark resonances
from octets vs $\mu$ have been obtained numerically (see Fig. 3).
They are avaluated around 230 MeV and 50 MeV,
correspondingly, i.e. this quantities are at least five times
smaller, than the mass and width of the scalar diquark resonance in
the color superconducting quark matter composed of $u$ and $d$
quarks \cite{eky1,eky2,he}.
 (Due to numerical difficulties, we were
not able to evaluate the parameters of the above mentioned massive
scalar- and pseudoscalar singlet diquark resonances of the CFL phase,
however, we guess that their masses are of the same order in
magnitude as the masses of the diquark octets.)

To get a more complete view about the diquark properties in the
framework of the NJL model (1), we have considered their masses in
the chirally broken phase of quark matter, too. It follows from
our analysis that i) at sufficiently strong interaction in the
diquark channel, i.e. at $G_2>H^{**}=1.5 G_1$, the existence of this
phase is prohibited in the framework of the NJL model (1), ii)
depending on the coupling constant $G_2$, scalar and pseudoscalar
diquarks have different properties in this phase. Indeed, at
$G_2<H^{*}$, where $H^*$ is given in (\ref{52}), both types of
diquarks are resonances. However, at $H^*<G_2<H^{**}$ the
pseudoscalar diquarks remain to be resonances, whereas scalar
diquarks are yet stable particles. As this takes place, there is a
splitting between the scalar diquark and scalar antidiquark masses
(see (\ref{53})), which is explained by the violation of the charge 
conjugation symmetry in the presence of a chemical potential. (Of
course, in the chirally broken phase all observable particles are
colorless, so one should expect that colored diquarks are
confined within baryons (see e.g. \cite{ebert}). Thus, one may
look at our investigation of the diquark masses in the chirally
broken phase as an indication of the existence of rather strong
quark-quark correlations inside baryons, which might help in a better
understanding of baryon dynamics.)

Finally, we have considered in the model (1) the masses of mesons 
which are stable particles in both phases. In the chirally broken
phase, i.e. at $\mu<\mu_c$, all nine pseudoscalar mesons are NG
bosons, whereas the nine scalar mesons have equal mass $\equiv 2M\sim
710$ MeV (for the parameter set of the model accepted in sec. II). In
the CFL phase, i.e. at $\mu>\mu_c$,
these nonet representations of mesons, reducible with respect to the
SU(3)$_{L+R+c}$ group, are decomposed into the octet and singlet
representations, each with its own mass. The reason for this
octet-singlet mass splitting of mesons is just  the color-flavor
locked symmetry breaking taking place at $\mu>\mu_c$. In the CFL
phase the masses of both types of mesons vary in the interval
300$\div$900 MeV, when $\mu$ varies from  330 MeV to 500 MeV (see
Figs 1,2). However, the mass splitting among the scalar and
pseudoscalar mesons occurs in different ways.
Indeed, as it is easily seen from Fig. 1, the mass of the scalar
octet mesons is larger than the mass of the scalar singlet meson,
whereas for pseudoscalar mesons the opposite situation takes place 
(see Fig. 2).

\section*{Acknowledgments}

The authors are grateful to the referee for fruitful remarks
concerning the multiplet structure of massive diquarks in the CFL
phase. This work has been supported in part by 
DFG project 436 RUS 113/477 and RFBR grant 05-02-16699.

\appendix

\section{Some formulae}
\label{ApB}
The present Appendix contains some useful formulae employed in the
text.\\
\noindent
{\bf i) Determinant:}
\begin{eqnarray}
\det\left
(\begin{array}{cc}
A~, & B\\
C~, & D
\end{array}\right )=\det [-CB+CAC^{-1}D]=\det
[DA-DBD^{-1}C].
\label{B1}
\end{eqnarray}
{\bf ii) Inverse matrix:}
\begin{eqnarray}
\left (\begin{array}{cc}
A~, & B\\
C~, & D
\end{array}\right )^{-1}=\left (\begin{array}{cc}
C^{-1}DL~, & -N\\
-L~~,&\!\!\!\!\!\! B^{-1}AN
\end{array}\right )=\left (\begin{array}{cc}
\bar L~~~~, &\!\!\!\!\!\! -A^{-1}B\bar N\\
-D^{-1}C\bar L~, & \bar N
\end{array}\right ),
\label{B2}
\end{eqnarray}
where
\begin{eqnarray}
L=[AC^{-1}D -B]^{-1}~~,~~N=[DB^{-1}A -C]^{-1}~~,
~~\bar L=[A-BD^{-1}C]^{-1}~~,~~\bar N=[D-CA^{-1}B]^{-1}.
\label{B3}
\end{eqnarray}
{\bf iii) Variational derivatives:} Let $A,B$ are some operators in
the coordinate space with matrix elements $A(x,y)\equiv$$A(x-y)$ and
$B(x,y)\equiv B(x-y)$, respectively. Moreover, let $\sigma(x)$ and
$\phi (x)$ are some fields. Then,
\begin{eqnarray}
{\rm Tr}\{A\sigma B\phi\}\equiv\int dx dy dz du
A(x,z)\sigma(z)\delta(z-y)B(y,u)\phi (u)\delta(u-x)=
\int dxdy A(x,y)\sigma(y)B(y,x)\phi(x).
\label{B4}
\end{eqnarray}
It follows from (\ref{B4}) that
\begin{eqnarray}
\Gamma (x-y)\equiv\frac{\delta^2{\rm Tr}\{A\sigma
B\phi\}}{\delta\sigma(y)\delta\phi(x)}=
A(x,y)B(y,x)=A(x-y)B(y-x).
\label{B5}
\end{eqnarray}
{\bf iv) Fourier transformations:} For arbitrary function $F(z)$ it
is possible to define the
Fourier transformation $\overline{F}(p)$ by the relation
\begin{eqnarray}
\overline{F}(p)=\int d^4z
F(z)e^{ipz},
~~~\mbox{i.~e.}~~~~ F(z)=\int\frac{d^4p}{(2\pi)^4}
\overline{F}(p)e^{-ipz}.
\label{B6}
\end{eqnarray}
Taking these relations into account, one obtains from
(\ref{B5}) that 
\begin{eqnarray}
\overline{\Gamma}(p)=\int\frac{d^4q}{(2\pi)^4}
\overline{A}(q+p)\overline{B}(q),
\label{B7}
\end{eqnarray}
where $\overline{A}(q),\overline{B}(q)$ are Fourier
transformations of the functions $A(x)$ and $B(x)$, correspondingly.

\section{Quark propagator matrix}

In the Nambu--Gorkov representation the inverse quark propagator
matrix $S_0^{-1}$ is given in (\ref{9}). Using the techniques,
elaborated in \cite{bekvy,eky1,eky2,he,hashimoto}, it is
possible to obtain the following expressions for the matrix elements
of the quark propagator matrix 
$S_0\equiv\vect S_{11},S_{12}\\S_{21},S_{22}\evect$:
  \begin{eqnarray}
\label{121} &&S_{11}(x-y)=\int\!\frac
{d^4q}{(2\pi)^4}\,e^{-iq(x-y)}\left\{
\frac{q_0-E^+}{q_0^2-(E_{B\Delta}^+)^2}
\gamma^0\bar\Lambda_++
\frac{q_0+E^-}{q_0^2-(E^-_{B\Delta})^2}
\gamma^0\bar\Lambda_-\right\},\\
&&S_{12}(x-y)=-i\Delta B\!\int\!\frac{d^4q}{(2\pi)^4}e^{-iq(x-y)}
\left\{\frac{1}{q_0^2-(E_{B\Delta}^+)^2}\gamma^5\bar\Lambda_-+
\frac{1}{q_0^2-(E^-_{B\Delta})^2}\gamma^5\bar\Lambda_+\right\},
\label{122}\\
&&S_{21}(x-y)=-i\Delta^*
B\!\int\!\frac{d^4q}{(2\pi)^4}e^{-iq(x-y)}\left
\{\frac{1}{q_0^2-(E_{B\Delta}^+)^2}\gamma^5\bar\Lambda_++
\frac{1}{q_0^2-(E^-_{B\Delta})^2}\gamma^5\bar\Lambda_-\right\},
\label{123}\\
&&S_{22}(x-y)=\int\!\frac{d^4q}{(2\pi)^4}e^{-iq(x-y)}\left\{
\frac{q_0+E^+}{q_0^2-(E_{B\Delta}^+)^2}
\gamma^0\bar\Lambda_-+
\frac{q_0-E^-}{q_0^2-(E^-_{B\Delta})^2}
\gamma^0\bar\Lambda_+\right\},\label{124}
\end{eqnarray}
where $M=\sqrt{\frac 23}\sigma$, $\bar\Lambda_\pm=\frac
12(1\pm\frac{\gamma^0(\vec\gamma\vec q- M)}{E})$. Moreover,
$(E_{B\Delta}^\pm)^2=(E^\pm)^2+|\Delta|^2B^2$,
$E^\pm=E\pm\mu$, $E=\sqrt{\strut\vec q^2+M^2}$ and
$B=\sum_{A=2,5,7} \tau_A\lambda_A$. (In these and other
similar expressions, $q_0$ is a shorthand notation for
$q_0+i\varepsilon\cdot {\rm sgn}(q_0)$, where the limit
$\varepsilon\to 0_+$ must be taken at the end of all
calculations. This prescription correctly implements the role of
$\mu$ as the chemical potential and preserves the causality
of the theory.) It is clear from (\ref{121})-(\ref{124}) that all
color- and flavor dependences in the matrix elements
$S_{11},S_{12},S_{21}$ and $S_{22}$ arise only due to the matrix $B$.
It is a $9\times 9$ matrix in the nine-dimensional space $c\times f$
which is the direct production of color and flavor spaces.
Note, in the chirally broken quark matter phase, where $\Delta =0$,
$M\ne 0$, the expressions for the matrix elements
(\ref{121})-(\ref{124}) have a simpler form. Namely, it is clear that
in this phase $S_{12}(x-y)=S_{21}(x-y)=0$ and
  \begin{eqnarray}
\label{1210} &&S_{11}(x-y)={\bf 1_c}\times {\bf 1_f}\times\int\!\frac
{d^4q}{(2\pi)^4}\,e^{-iq(x-y)}\left\{
\frac{\gamma^0\bar\Lambda_+}{q_0+E^+}
+\frac{\gamma^0\bar\Lambda_-}{q_0-E^-}
\right\},\\
&&S_{22}(x-y)={\bf 1_c}\times {\bf 1_f}\times
\int\!\frac{d^4q}{(2\pi)^4}e^{-iq(x-y)}\left\{
\frac{\gamma^0\bar\Lambda_-}{q_0-E^+}+
\frac{\gamma^0\bar\Lambda_+}{q_0+E^-}
\right\},\label{1240}
\end{eqnarray}
i.e. the matrix elements (\ref{1210})-(\ref{1240}) are
proportional to the unit matrices both in the color and flavor
spaces.

\end{document}